\documentclass[conference]{IEEEtran}

\usepackage[textsize=large]{todonotes}
\usepackage{booktabs}
\usepackage{multirow}
\usepackage{wrapfig}
\usepackage{graphicx}
\usepackage{graphics}
\usepackage{subcaption}
\usepackage{placeins}
\usepackage[rawfloats=true]{floatrow} 
\usepackage[numbers]{natbib}
\usepackage{url}
\usepackage{adjustbox}
\usepackage{amsmath}
\usepackage{array}

\begin{document}

\title{N-TORC: Native Tensor Optimizer for\\Real-time Constraints}
\author{
Suyash Vardhan Singh\textsuperscript{1}, Iftakhar Ahmad\textsuperscript{1}, David Andrews\textsuperscript{2},\\Miaoqing Huang\textsuperscript{2}, Austin R.J. Downey\textsuperscript{1}, Jason D. Bakos\textsuperscript{1}\\
\textsuperscript{1}University of South Carolina, Columbia, South Carolina, USA\\
\textsuperscript{2}University of Arkansas, Fayetteville, Arkansas, USA\\
\{ss121@email.sc.edu, iahmad@email.sc.edu, dandrews@uark.edu,\\mqhuang@uark.edu, austindowney@sc.edu, jbakos@cse.sc.edu\}
}


%


\maketitle


\begin{abstract}
Compared to overlay-based tensor architectures like VTA or Gemmini, compilers that directly translate machine learning models into a dataflow architecture as HLS code, such as HLS4ML and FINN, generally can achieve lower latency by generating customized matrix-vector multipliers and memory structures tailored to the specific fundamental tensor operations required by each layer. However, this approach has significant drawbacks: the compilation process is highly time-consuming and the resulting deployments have unpredictable area and latency, making it impractical to constrain the latency while simultaneously minimizing area. Currently, no existing methods address this type of optimization. In this paper, we present N-TORC (Native Tensor Optimizer for Real-Time Constraints), a novel approach that utilizes data-driven performance and resource models to optimize individual layers of a dataflow architecture. When combined with model hyperparameter optimization, N-TORC can quickly generate architectures that satisfy latency constraints while simultaneously optimizing for both accuracy and resource cost (i.e. offering a set of optimal trade-offs between cost and accuracy). To demonstrate its effectiveness, we applied this framework to a cyber-physical application, DROPBEAR (Dynamic Reproduction of Projectiles in Ballistic Environments for Advanced Research). N-TORC's HLS4ML performance and resource models achieve higher accuracy than prior efforts, and its Mixed Integer Program (MIP)-based solver generates equivalent solutions to a stochastic search in 1000X less time.
\end{abstract}


%
\IEEEpeerreviewmaketitle

\section{Introduction}


Interest is growing in developing high-rate cyber-physical systems that must make control decisions within one millisecond. These systems often rely on predictive models of physical phenomena. However, the tight time constraints make physics-based models impractical. Lightweight machine learning models offer a promising alternative, surrogate model, providing sufficient accuracy while ensuring deterministic, sub-millisecond latency \cite{satme2022progress,kabir2023accelerating,vereen2023optimal,ogunniyi2023real,Downey2021Dataset2Dropbear}.


An early and well-known attempt for this approach was the JetDNN jet substructure classifier designed for the Large Hadron Collider \cite{que2023metaml,wielgosz2017using,egan2017long}. JetDNN is a small network model, consisting of only 4,256 parameters arranged in dense layers only, but its performance has been extensively tuned using quantization and compression techniques \cite{que2023metaml} and conversion to a LogicNet \cite{9221584}. However, larger, more complex physics-based models that contain a mixture of convolutional layers and LSTM layers create substantially longer design space and compilation times, making these techniques less practical \cite{YAN2021107960,vibration3030016,10.1117/12.3010900}.

Neural network-to-High-Level Synthesis (HLS) compilation flows, such as HLS4ML, generate matrix multipliers with physical dimensions that evenly divide the tensor operations required by each layer. This approach achieves optimal latency and supports the use of different hardware precisions for each layer. However, it results in low hardware utilization, because only the matrix multiplier associated with one layer can be active at a time. Despite this limitation, the physical size of the matrix multipliers for each layer can be independently adjusted, enabling fine-grained control over trade-offs between time and resource usage. Exploiting this flexibility remains an open challenge, as accurately predicting how layer configurations affect resource requirements is difficult.


In this paper, we describe a framework for simultaneous neural architecture search and deployment optimization. Our neural architecture search optimizes against both model accuracy and workload, the latter of which serves as an approximation of the cost and performance of the deployed model. This produces a Pareto optimal set, and for each member of this set we optimize its deployment using a Mixed Integer Program (MIP)-based solver to assign the native size of the matrix multiplier assigned to each network layer on the FPGA. The solver uses performance and cost models trained from HLS4ML for a specific target FPGA to predict the cost and latency of each layer. 




Our accuracy results are evaluated using the DROPBEAR (Dynamic Reproduction of Projectiles in Ballistic Environments for Advanced Research) dataset, described below \cite{Downey2021Dataset2Dropbear}. 
The source code for our performance models, datasets, deployment optimizer, and network optimizer is publicly available via a Github repo \cite{projectrepo}. This paper describes three main contributions:

\begin{enumerate}
    \item a multi-objective Bayesian hyperparameter search,
    \item an MIP-based optimizer that optimizes the dimensions of each physical functional unit to constrain the latency while minimizing resource cost, and
    \item a set of linear performance and cost models for three types of HLS4ML layers (convolution, LSTM, and dense), which is integrated into the MIP solver.
\end{enumerate}

\section{Background}



\begin{figure}[!t]
    \centering
    \includegraphics[width=3in]{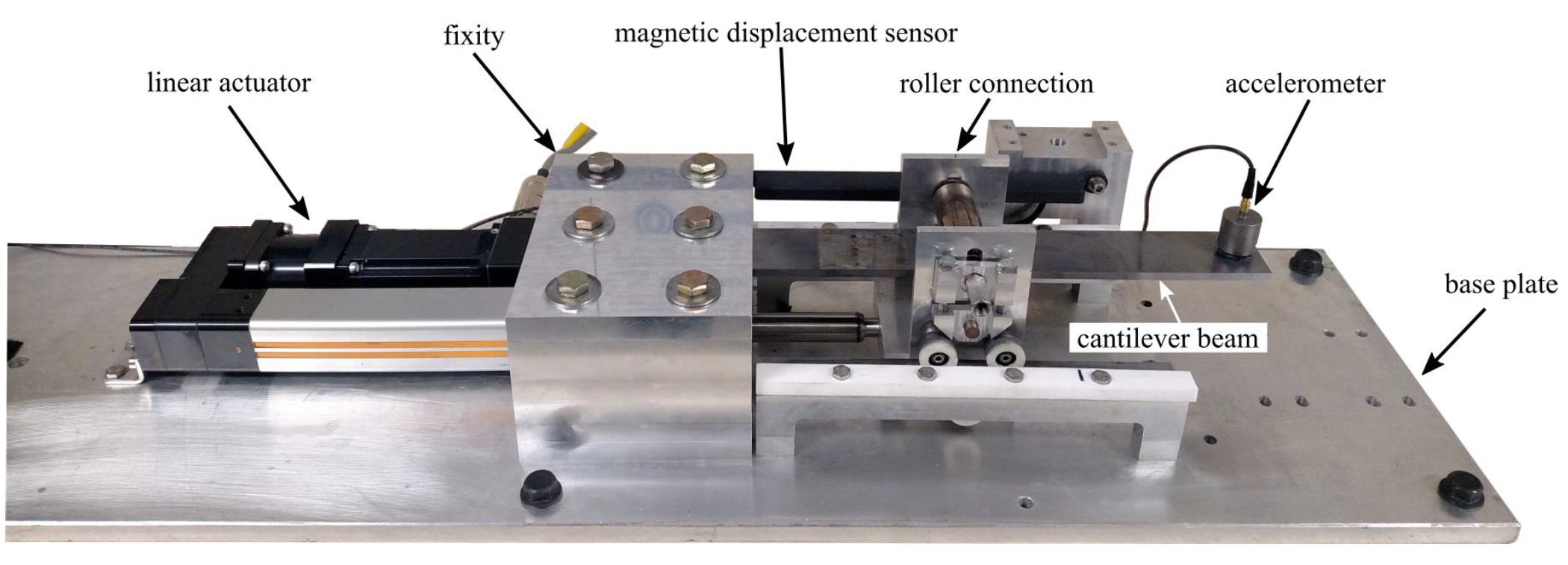}
    \caption{The DROPEAR experimental setup which consists of a cantilever beam with a movable roller and an accelerometer mounted on the bottom of the beam.}
    \label{fig:dropbear}
\end{figure}

The Dynamic Reproduction of Projectiles in Ballistic Environments for Advanced Research (DROPBEAR) \cite{Nelson2022GeneratedDatasetsDynamic} was used to generate the experimental data used in this work. DROPBEAR is meant to serve as an exemplar for real-time, high-rate machine learning applications \cite{Dodson2021HighRateStructural,panahi2022high}. The DROPBEAR testbed is presented in Fig. \ref{fig:dropbear}. It is a cantilever beam with a controllable roller support that moves to alter the boundary condition of the beam, which is designed to produce a repeatable change in the structural state of the system that can ideally be inferred in real time from its vibration measured from an accelerometer.

\begin{figure}[!t]
    \centering
    \includegraphics[width=2.5in]{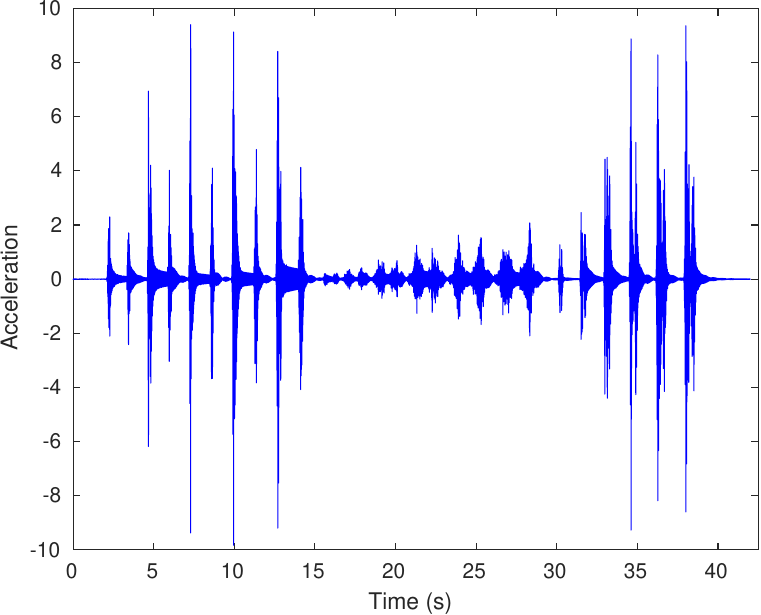}
    \caption{DROPEAR acceleration data, which results from the roller movements but is treated as an input into a model that predicts the roller location given the acceleration signal.}
    \label{fig:acceleration_data}
\end{figure}

\begin{figure}[!t]
    \centering
    \includegraphics[width=2.5in]{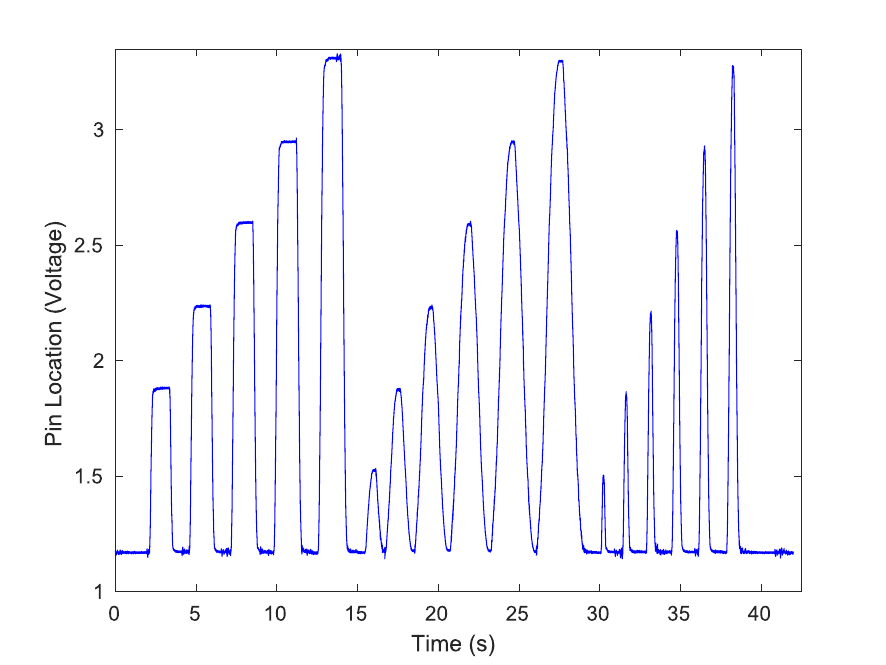}
    \caption{DROPEAR roller position, which moves to simulate a moving boundary condition for the cantilever beam. Essentially it sets the root of the cantilever beam.}
    \label{fig:pin_location_data}
\end{figure}

As shown in Fig. \ref{fig:pin_location_data}, the roller followed a movement pattern ranging from 58 mm to 141 mm. The beam is self-excited by the movements of the roller, and therefore, no external input is required. 
The maximum roller speed was limited by the experimental setup to 250 mm/s. An example time series acceleration response of the beam is presented in Fig. \ref{fig:acceleration_data} and the corresponding roller location is shown in Fig. \ref{fig:pin_location_data}. The data used in this work are available through a public repository \cite{Downey2021Dataset2Dropbear}.

\subsection{Targeted Layer Types}
\label{sec:layer_types}

The objective of the Working Group on High-rate Structural Monitoring, Damage Detection, Prognostics, and Reactions \cite{hrwg} is to develop models that predict the state of a structure given its vibration data and to do so at high rate ($< 1$~ms). Such models would enable new approaches for monitoring structural health (e.g. detecting damage on the wing of a hypersonic vehicle or controlling a smart airbag system that reacts to the dynamics of a collision) and designing smart structures (e.g. the design of morphing wings without traditional control surfaces).
The goal of DROPBEAR is to develop a model that takes as input a time series vibration signal and produces the corresponding roller location with its inference time bounded to $200~\mu\text{s}$, due to the dataset's 5 KHz sample rate. Models may leverage Takens Embedding Theorem \cite{10.1007/BFb0091924}, which states that the state of a chaotic dynamical system can be estimated from a sequence of observations taken at fixed time delays, i.e. samples from time $t, t-\tau, t-2\tau, t-3\tau, ...$ where $t$ is the current time and $\tau$ is the time delay.

The challenge presented by DROPBEAR is an inverse physics problem, in which the vibration signal, caused by the roller movement, is used to predict the roller position. In other words, in the physical system, the roller movement is the cause and the vibration is the effect, while in the modeling problem, the vibration signal is the input and the roller position is the output. Prior efforts to design models for DROPBEAR consisted of physics-based models \cite{10.1117/12.2613208,OGUNNIYI2023110318,10224187} and data-driven models. The data-driven models were regression neural networks consisting only of LSTM layers with a single dense output layer \cite{satme2022progress,kabir2023accelerating}. In this paper, we consider more sophisticated models by adding front-end 1D convolution layers and additional backend dense layers with LSTM layers in between.



Our network accepts a vector consisting of $n$ samples, which are sent into a sequential network consisting of a 1D convolution stage comprised of convolution + ReLU + max pooling blocks, followed by an LSTM layer stage, followed by a dense stage. We generate and train a sequence of networks in this pattern, independently setting the number of inputs $n$, the size of each layer, and the number of each type of layer. For each network generated and trained, we compute its accuracy as root mean square error (RMSE) and its total workload in total number of multiplies required for the forward pass. 1D convolution layers perform $s \times k \times f_1 \times f_2$ multiplies, where $s$ = the sequence length, $k$ = the filter kernel size, $f_1$ = the input feature length and $f_2$ = the output feature length. LSTM layers perform $(s \times f + u) \times (4 \times u)$ multiplies, where $s$ = the input sequence length, $f$ = the input feature length, and $u$ = the number of LSTM units. Dense layers perform $f \times n$ multiplies, where $f$ = the number of input features (which may need to be flattened from a multidimensional input tensor) and $n$ = the number of neurons.

\subsection{HLS4ML}

HLS4ML \cite{duarte2018fast} was originally developed in 2018 by a team connected with researchers at FermiLab and CERN with the objective of developing a sub-microsecond latency FPGA-based neural network to serve as the first stage of a filter for data sampled in the Large Hadron Collider. HLS4ML comprises a library of handwritten parameterized neural network layers that are automatically instantiated and interconnected based on a network model described in Keras.

Coincident with the original development of HLS4ML, two neural networks were trained to classify jet substructures detected within the Large Hadron Collider. These networks were small, consisting of four dense layers and requiring a total of $4,256$ multiplies, but were purposely kept small to allow full parallelization on an FPGA. In this scenario, ``full parallelization'' implies that one hardware multiplier is instantiated in hardware and assigned to every weight in the neural network.

\subsubsection{HLS4ML Reuse Factor}

Unlike JetDNN, not all networks deployed with HLS4ML need to be fully parallelized, as the resource cost and compilation time for fully parallelized networks can be prohibitive for larger networks. HLS4ML parallelization is focused on the innermost two loops that form the core of each layer type. Of these, the outer loop iterates \textbf{n\_in} times and the inner loop iterates \textbf{n\_out} times, as named in the HLS4ML source code. In other words, each layer performs a matrix-vector multiply of size \textbf{n\_in} $\times$ \textbf{n\_out}, regardless of the physical size of the multiplier generated. For a dense layer, n\_in = number of input features, n\_out = number of neurons. For a Conv1D layer, n\_in = number of input channels multiplied by the filter kernel size, n\_out = number of output channels. For an LSTM layer, n\_in = number of input features, n\_out = number of LSTM units multiplied by 4. 

In HLS4ML, the desired level of parallelization for each layer is expressed through a parameter called ``reuse factor.'' Reuse factor is a deployment parameter assigned to layer and determines the number of times a hardware multiplier is used when executing the innermost two loops for a layer's forward pass. A reuse factor of $R$ means that each hardware multiplier is used to perform $R$ of the multiplies required in the innermost two loops, meaning that the number of instantiated hardware multipliers is only $1/R$ that of a fully parallelized ($R=1$) realization of the same layer. The reuse factor must evenly divide \textbf{n\_in} $\times$ \textbf{n\_out}. The number of physical multipliers instantiated is known as the ``block factor'', shown in Eq. \ref{eq:block_factor}.

\begin{equation}
\text{block\_factor} =
\left\lceil
    \frac{\text{n\_in} \cdot \text{n\_out}}{R}
\right\rceil
\label{eq:block_factor}
\end{equation}

\begin{figure}[!t]
    \centering
    \includegraphics[width=.5\textwidth]{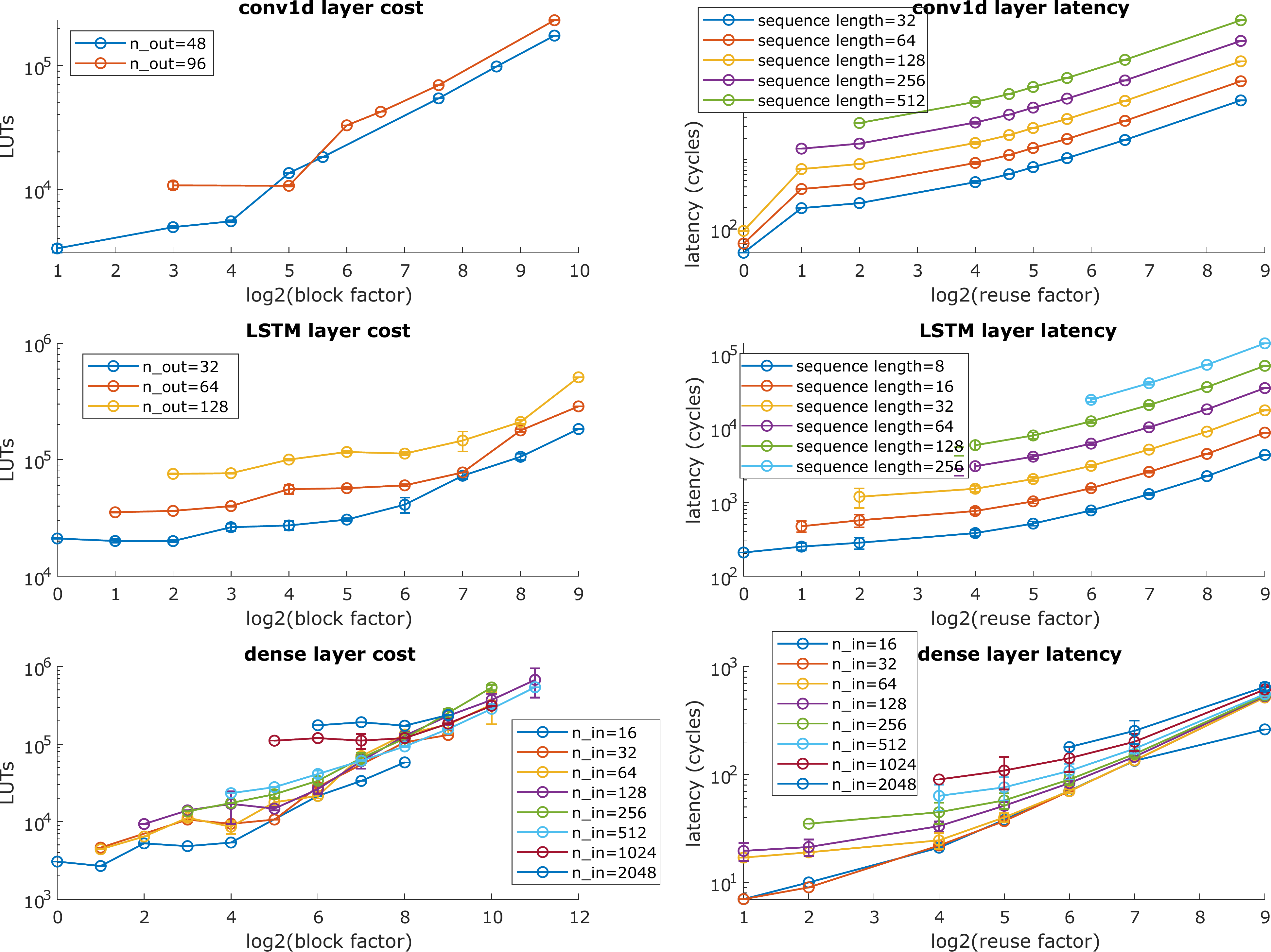}
    \caption{LUT cost and latency when scaling size of hardware GEMV unit (as block factor = number of scalar multipliers) for three types of HLS4ML layers. Note that each data point represents a set of observations, as many layer configurations map to the same independent variables shown in the plots (e.g. various combinations of hyperparameters that imply the same values of \textbf{n\_in} and \textbf{n\_out}); error bars indicate standard deviation.}
    \label{fig:cost_plots}
\end{figure}

We have observed that a layer's resource cost in LUTs, FFs, and DSPs is a function of its block factor and either the number of iterations of the inner loop, \textbf{n\_out}, or the number of iterations of its outer loop, \textbf{n\_in}, depending on the layer type. On the other hand, a layer's latency is a function of the layer's reuse factor and its sequence length. The sequence length comprises the number of trips through a sequential loop that encloses both the \textbf{n\_in} and \textbf{n\_out} loops, and whose trip count is set to the number of outputs per channel from convolution layers and is carried through downstream LSTM layers. Dense layers do not accept a sequence length, so the embedding dimension and sequence length are flattened when fed into a dense layer, becoming the number of outer loop iterations, \textbf{n\_in}.

LUT cost and latency for conv1d, LSTM, and dense layers are shown in Fig. \ref{fig:cost_plots}. Note that each data point is computed as the mean of observed resource/latency values for a set of synthetically-generated layers that fall into the category defined by the corresponding block factor and \textbf{n\_out}/\textbf{n\_in} (for resource) or the reuse factor and the sequence length (for latency). Error bars are shown for each data point. As shown in the figure, latency is reasonably predictable, whereas resource cost is less predictable.





\subsubsection{Scale of Deployed Networks}

The networks targeted in this paper accept up to 512 inputs and contain zero to five groups of 1D convolution + activation + pooling layers having up to 256 feature maps each (with each layer requiring no greater than 100,663,296 multiplies), followed by zero to three Long Short Term Memory (LSTM) layers having up to 425 units each (with each requiring no greater than 223,544,900 multiplies), followed by one to five dense layers having up to 512 neurons each (with each layer requiring no greater than 111,411,200 multiplies). Our largest possible network would comprise 435,619,396 multiplies, although in practice, the hyperparameter optimization used in this work generally yields networks that are less than 700,000 multiplies and the Pareto optimal set of networks generally requires 10,000 to 40,000 multiplies, indicating that networks larger than this do not deliver sufficiently higher accuracy to justify their cost.

\section{Hyperparameter Optimization}

Our proposed framework consists of a two-phase neural architecture search. First, the framework selects an optimal set of hyperparameters to maximize the accuracy of the model while minimizing the workload of the model. Second, each of the optimal models found is deployed by optimizing the reuse factor for each layer in order to constrain end-to-end latency and minimize total model resource cost.

Our model hyperparameter search is performed using the Optuna framework \cite{optuna_2019} V 4.0.0. We used the Bayesian Optimization multi-objective sampling strategy in PyTorch (BoTorch) \cite{balandat2020botorch} from the ``optuna-integration module'', which uses a Quasi-Monte Carlo acquisition function. Our objective function is to minimize root-mean-square-error (RMSE) of the validation set and implied network workload (number of multiplies).

Each network is trained using Dataset 8 from the High-rate Structural Monitoring, Damage Detection, Prognostics, and Reactions Working Group \cite{Downey2021Dataset2Dropbear}. This dataset consists of 150 separate experimental runs, each composed of a vibration signal and the corresponding roller position signal.

\subsection{Dataset}
\label{dataset_section}

These datasets are grouped into three experimental types, each defined by the pattern used to stimulate roller movement. Each dataset samples both acceleration and roller position at 5 KHz ($200~\mu\text{s}$ per sample).

\begin{enumerate}
    \item \textbf{Standard Index Set (20 datasets):} This roller movement pattern resembles that shown in Fig. \ref{fig:pin_location_data} in which the roller performs a series of square waves of increasing magnitude, followed by the function $abs(sin(x))$ of increasing magnitude, followed by the function $min(sin(x),0)$ of increasing magnitude.
    
    \item \textbf{Random Dwell (100 datasets):} This roller movement pattern moves to the roller to random locations at fixed intervals.
    
    \item \textbf{Slow Positional Displacement (30 datasets):} This roller movement pattern advances the roller in increments until it reaches its maximum value and then retracts the roller in increments until it reaches its starting point. After each change in roller location, the roller pauses for a fixed amount of time.
\end{enumerate}


We randomly selected 15 datasets (12 for training and 3 for testing) from each category. In total, we use 36 datasets for training and 9 for testing, which we refer to as ``Test Dataset 1''. This data is also shuffled, and the training data is split into a 70-30\% training-validation split. We refer to the validation portion of the training  as ``Test Dataset 2''.

\subsection{Network Hyperparameter Search}



\begin{figure}[!t]
    \centering
    \includegraphics[width=3in]{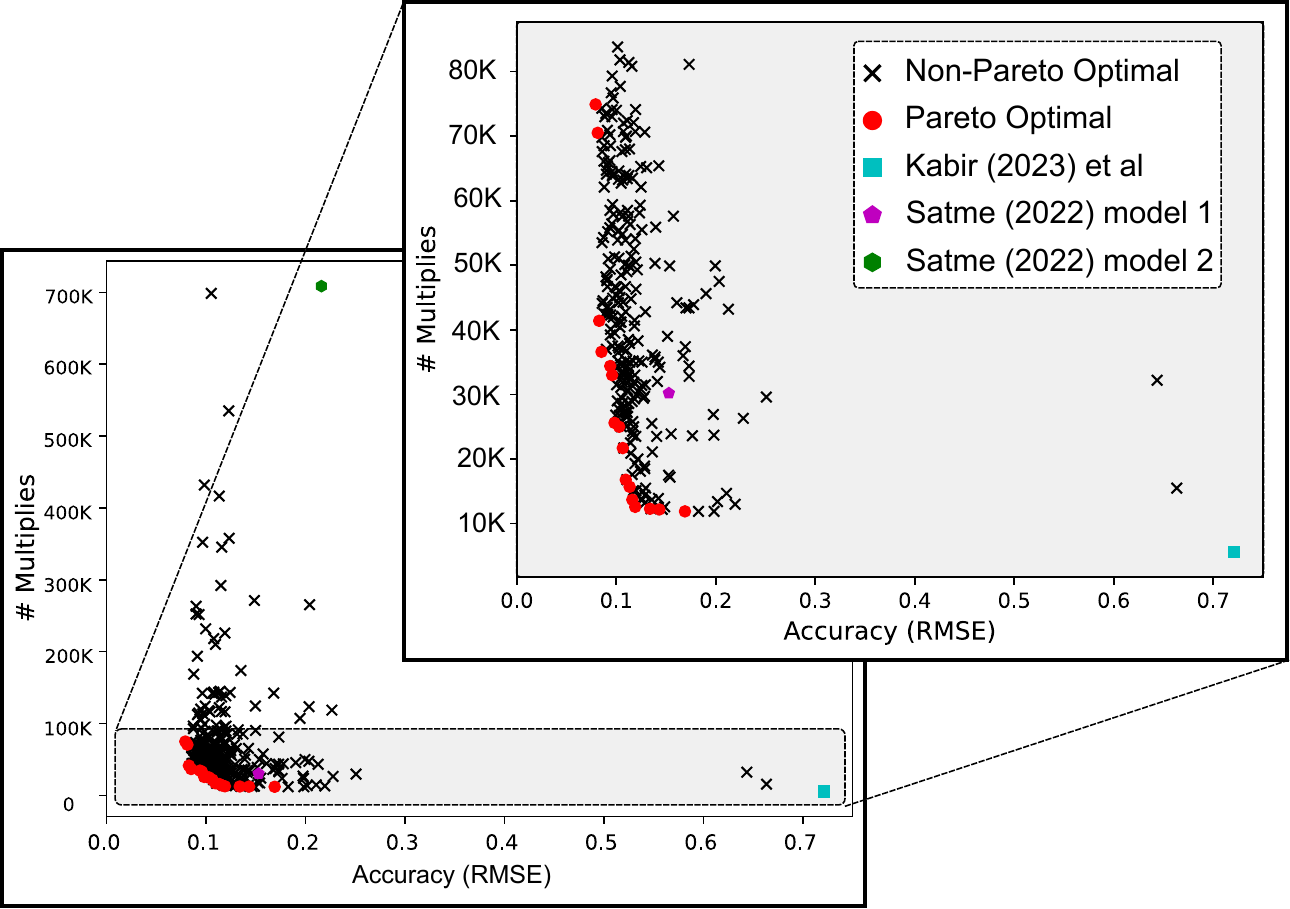}
    \caption{Pareto optimal model configurations for accuracy and cost. Included are the positions of Satme et al. network 1 and 2 (purple and green dots) \cite{satme2022progress} and Kabir et al. (cyan square) \cite{kabir2023accelerating}.}
    \label{fig:pareto}
\end{figure}

Fig. \ref{fig:pareto} is a Pareto optimal plot showing the results from a hyperparameter search. The Pareto optimal networks are shown as red dots and non-optimal networks are shown as black dots. Also shown are points from networks designed for DROPBEAR in previous work; networks designed by Satme et el. \cite{satme2022progress} are shown as purple and green dots, and the networks designed by Kabir et al. \cite{kabir2023accelerating} are shown as a cyan square. All three of these models were re-trained with the same training data and evaluated with the same test data as our networks. Only one network designed by Satme (network 1) is situated near the Pareto front. We used \textit{Test Dataset 2} described in Sec. \ref{dataset_section} to calculate the RMSE shown in Fig. \ref{fig:pareto}.



\section{Automated Model Deployment}




HLS4ML requires a ``reuse factor'' for each layer. The reuse factor determines the latency and the number of required LUTs, FFs, DSPs, and BRAMs for the layer. These values can only be determined by compiling the model, which for most networks can require several CPU-days to perform. For this reason, it is necessary to develop models that can predict the resource cost and latency of a layer given the associated parameters, which include the layer type, its tensor dimensions, its size, and its reuse factor.

To estimate latency and resource cost for each of the target layer types, we trained a random forest regression model from Scikit-Learn \cite{scikit-learn},
The total latency and cost for a given network can be obtained by evaluating these models for each layer comprising the network and summing up the outputs.

In order to train our performance and cost models, we synthesized a series of networks corresponding to nearly every permutation of the following set of parameters, which determine the depth of the network and size of each layer:

\begin{itemize}
    \item \textit{Feature inputs:} 128, 256, 512
    \item \textit{Number of 1D convolutional layers:} 1, 2, 4
    \item \textit{Output channels/convolutional layer:} 16 and 32
    \item \textit{Number of LSTM layers:} 0, 1, 2
    \item \textit{LSTM units/LSTM layer:} 8, 16, 32
    \item \textit{Number of dense layers:} 1, 2, 4
    \item \textit{Number of neurons/dense layer:} 16, 32, 64
    \item \textit{Raw reuse factors (corrected as needed for each layer):} 1, 2, 4, 16, 32, 64, 128, 512
\end{itemize}

Each network is synthesized using Vivado HLS 2019.1 to obtain resource and latency estimates for each layer, obtained by extracting the relevant data from the report files generated by the HLS compiler. Each layer is identified as its layer type and the following features:

\begin{itemize}
    \item input size as a 2D tensor (e.g. number of features/embedding dimension and sequence length),
    \item layer size (e.g. number of output feature maps, LSTM units, dense neurons), and
    \item reuse factor.
\end{itemize}

For each synthesis run, we extracted the resultant cost of each layer in LUTs, BRAMs, DSPs, and flip-flops, and the reported latency in cycles for the target FPGA, which in this case is the Zynq Ultrascale+ ZU7EV. In total, we synthesized 11,851 networks. The target clock for the synthesized designs was 250 MHz and the precision used was 16 total bits and 8 fractional bits. Since the input size of each hidden layer is determined by its predecessor layer, many generated layers have the same features. All samples having the same features are averaged into a single observation. The total number of unique layers obtained was: 5,962 dense layers, 496 LSTM layers, and 4,195 1D convolutional layers.

\subsection{Model Accuracy}

Table \ref{tab:validation_metrics} reports the testing accuracy of these models reported when using 80\% of our compiled results as training data and 20\% as testing data. The evaluation reports $R^2$ score, Mean Absolute Error (MAE) percentage, and Root Mean Square Error (RMSE) percentage.

The $R^2$ scores indicate strong model performance, but some MAE and RMSE percentages exhibit variability, particularly for the LSTM layers. For example, the BRAM metric for LSTM has a MAE of 11.98\%  and a RMSE of 23.37\%, suggesting the presence of hidden variables or stochastic behavior in the compiler. In contrast, the MAE and RMSE percentages for convolutional and dense layers are relatively lower, demonstrating a higher prediction accuracy.
 
\begin{table}[]
\centering
\resizebox{\columnwidth}{!}{
\begin{tabular}{|l|l|c|c|c|c|}

\hline
\textbf{Layer}                 & \textbf{Metric}               & \textbf{$R^2$ Score} & \textbf{MAPE} & \textbf{RMSE \%} & \textbf{Value Range} \\ \hline
\multirow{5}{*}{Convolutional} & BRAM    & 0.9976               & 0.44        & 6.76          & 0 - 342 \\ \cline{2-6} 
                               & LUT     & 0.9988               & 2.35        & 3.95          & 2121.82 - 231963 \\ \cline{2-6} 
                               & FF      & 0.9995               & 0.60        & 1.84          & 1042 - 75576 \\ \cline{2-6} 
                               & DSP     & 0.9979               & 1.21        & 6.86          & 1 - 768 \\ \cline{2-6} 
                               & Latency & 0.9999               & 0.09        & 0.71          & 45 - 101910 \\ \hline
\multirow{5}{*}{LSTM}          & BRAM    & 0.9371               & 11.98       & 23.37         & 16 - 489 \\ \cline{2-6} 
                               & LUT     & 0.9800               & 1.36        & 11.16         & 18580.714 - 286843 \\ \cline{2-6} 
                               & FF      & 0.9826               & 1.23        & 10.06         & 7680.33 - 87131 \\ \cline{2-6} 
                               & DSP     & 0.9780               & 1.65        & 15.54         & 26 - 1072 \\ \cline{2-6} 
                               & Latency & 0.9988               & 2.59        & 6.00          & 209 - 140545 \\ \hline
\multirow{5}{*}{Dense}         & BRAM    & 0.9954               & 0.13        & 11.48         & 0 - 910 \\ \cline{2-6} 
                               & LUT     & 0.9921               & 0.14        & 15.17         & 1203 - 1079840 \\ \cline{2-6} 
                               & FF      & 0.9989               & 0.09        & 4.89          & 1269 - 206076 \\ \cline{2-6} 
                               & DSP     & 0.9956               & 0.12        & 13.54         & 1 - 2048 \\ \cline{2-6} 
                               & Latency & 0.9931               & 4.20        & 10.18         & 7 - 793 \\ \hline
\end{tabular}}
\caption{Validation metrics for convolutional, LSTM, and dense layers across different resource and latency metrics, using MAE and RMSE percentages for accuracy over the range.}
\label{tab:validation_metrics}
\end{table}

\begin{table*}[]
\begin{center}

\begin{tabular}{|l|c|c|c|c|c|c|}

\hline
\textbf{Metric} & \textbf{\begin{tabular}[c]{@{}c@{}}Best MAPE\\ (\cite{wu2022high})\end{tabular}} & \textbf{\begin{tabular}[c]{@{}c@{}}Best MAPE\\ (This work)\end{tabular}} & \textbf{\begin{tabular}[c]{@{}c@{}}Median MAPE\\ (\cite{wu2022high})\end{tabular}} & \textbf{\begin{tabular}[c]{@{}c@{}}Median MAPE\\ (This work)\end{tabular}} & \textbf{\begin{tabular}[c]{@{}c@{}}Worst MAPE\\ (\cite{wu2022high})\end{tabular}} & \textbf{\begin{tabular}[c]{@{}c@{}}Worst MAPE\\ (This work)\end{tabular}} \\ \hline
DSP             & 8.95                                                                             & 1.21                                                                     & 10.98                                                                              & 0.12                                                                       & 15.03                                                                             & 2.59                                                                     \\ \hline
LUT             & 4.02                                                                             & 0.14                                                                     & 10.27                                                                              & 1.36                                                                       & 26.33                                                                             & 2.35                                                                     \\ \hline
FF              & 5.78                                                                             & 0.09                                                                     & 11.22                                                                              & 0.60                                                                       & 25.52                                                                             & 1.23                                                                     \\ \hline
Latency         & 4.91                                                                             & 0.09                                                                     & 5.81                                                                               & 2.59                                                                       & 8.72                                                                              & 4.20                                                                     \\ \hline
BRAM            & N/A                                                                              & 0.13                                                                     & N/A                                                                                & 1.58                                                                              & N/A                                                                               & 11.98                                                                     \\ \hline
\end{tabular}
\end{center}
\caption{Comparison of MAPE values between Wu et al.'s approach \cite{wu2022high} and our approach for different resource and latency metrics, considering best, median, and worst values.}
\label{tab:comparison_metrics}
\end{table*}

In order to illustrate the challenge of estimating resource and latency associated with synthesized HLS code, we compare our proposed estimation method against that of Wu et al. \cite{wu2022high}, which makes its predictions using a Graph Neural Network-based model that accepts HLS intermediate code as its input. The comparison shown in Table \ref{tab:comparison_metrics} compares the best, median, and worst mean average percentage error (MAPE\%) across a set of benchmark kernels tested by Wu et al. versus results from the three HLS4ML layer types that we tested using our proposed method against our test set. Our models achieve lower MAPE percentages for both best and median cases, indicating higher accuracy, potentially owed to the fact that our proposed model is trained and used only for HLS4ML layers as opposed to being applicable to any HLS code.

\subsection{Reuse Factor Optimizer}

\begin{figure}[!t]
    \centering
    \includegraphics[width=0.3\textwidth]{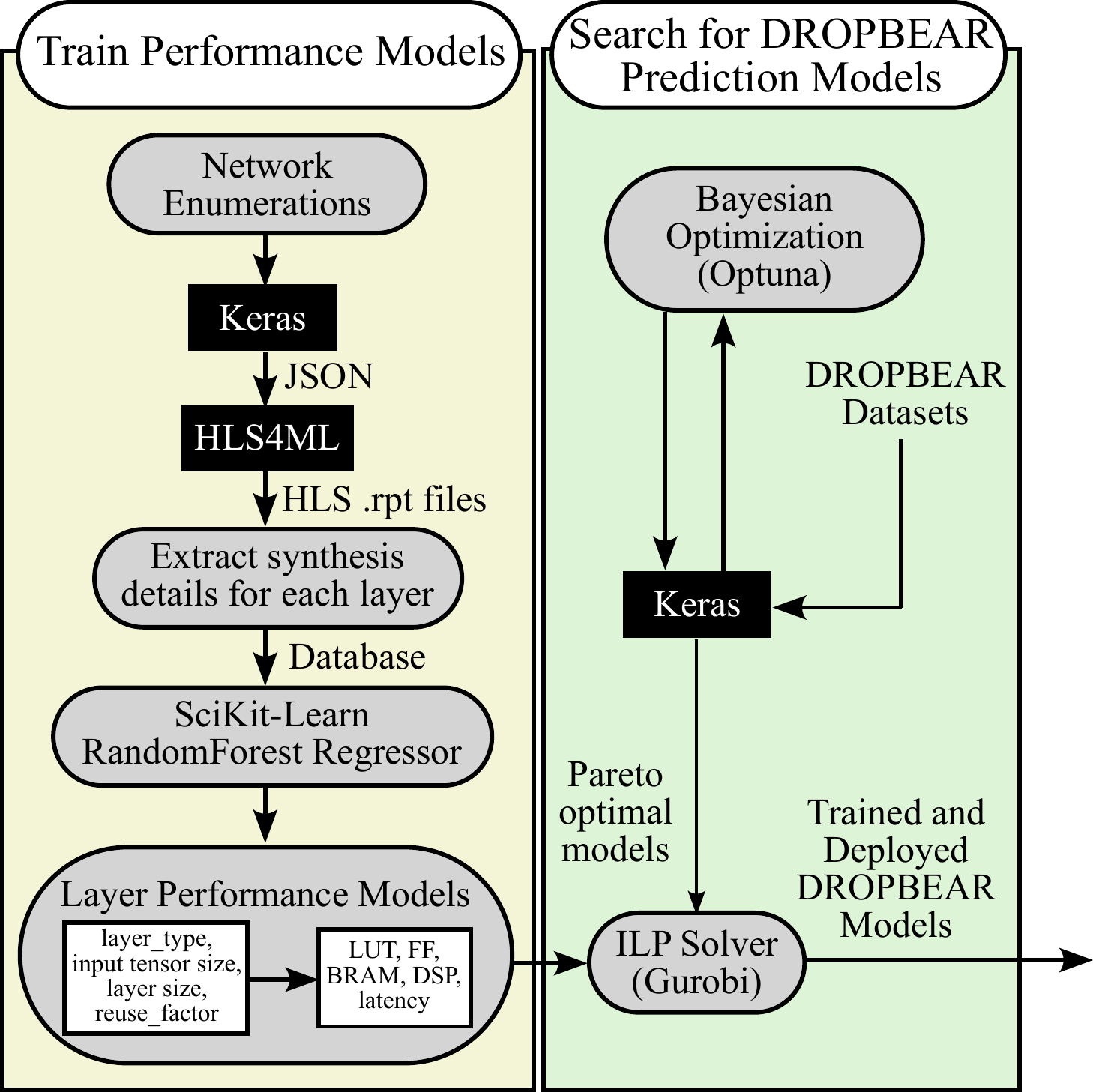} 
    \caption{Overview of N-TORC Framework.}
    \label{fig:tool_flow}
\end{figure}

To determine the optimal reuse factors for a deployed model, we use the Gurobi Mixed Integer Programming (MIP) solver \cite{gurobi} to solve the following problem:

\[
\begin{aligned}
    \text{Minimize:} & \quad \sum_{i \in \text{layers}} \big( \widehat{\text{LUTS}_i} + \widehat{\text{FF}_i} + \widehat{\text{BRAM}_i} + \widehat{\text{DSP}_i} \big) \\
    \text{Subject to:} & \quad \sum_{i \in \text{layers}} \widehat{\text{latency}_i} \leq 50000
\end{aligned}
\]

where $\widehat{\text{LUTS}_i},  \widehat{\text{FF}_i}, \widehat{\text{BRAM}_i}, \widehat{\text{DSP}_i}, and \
 \widehat{\text{latency}_i}$, are the estimations of the LUTs, FFs, BRAMs, and DSPs for layer $i$, as given by the cost and performance models.

For a given layer type (e.g. conv1d, LSTM, dense) and a set of associated layer parameters (e.g. input tensor size, layer size), Gurobi automatically converts the random forest model into a linear model. Note that although the random forest model is trained with multiple inputs, when deployed in our optimizer we set all inputs to constants except for the reuse factor. This allows the model to collapse into a linear expression, allowing it to be used for the mixed integer solver.


\begin{figure}[!t]
    \centering
    \includegraphics[width=3in]{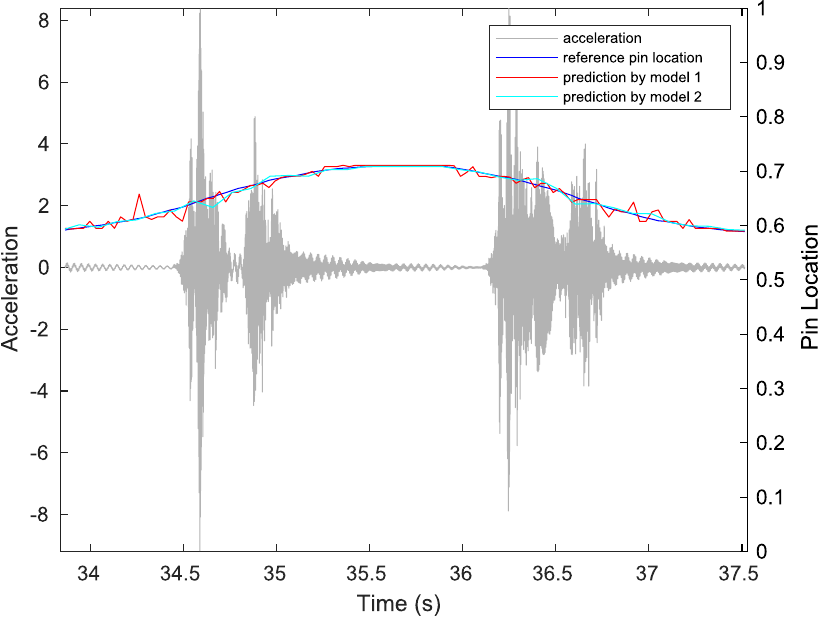}
    \caption{A zoomed-in portion of DROPBEAR, showing vibration signal input (gray), ground truth roller position (blue), and the results of two trained models with RMSE ~= 0.07 (cyan) and RMSE ~= 0.119 (red).}
    \label{fig:dropbear_accuracy}
\end{figure}

\section{N-TORC Framework}

Fig. \ref{fig:tool_flow} summarizes the complete N-TORC toolflow \cite{suyash}. The left side depicts the process by which networks are generated and compiled in order to create a database of ground truth data with which to train the performance and resource models.

Each generated network is converted into Keras format and run through the HLS4ML flow. The resulting latency and resource requirements for each layer are extracted from the report files. Using this data, we train six random forest regression models for each layer type against resources and latency (conv1d, LSTM, and dense) using an 80-20 train/test mix.

The right side of the figure shows the hyperparameter optimization, which searches for network configurations that achieve Pareto optimal accuracy and cost as described in Sec. \ref{sec:pareto}. For each of these, N-TORC generates a Gurobi MIP program that includes the features (input tensor and size) for each layer, combined with the trained HLS4ML performance and accuracy models for each layer type on the target FPGA. Executing this program sets the reuse factor of each layer to meet the real-time latency constraint while minimizing resource cost.

\section{Experimental Results}

\begin{table*}[]
\begin{center}
\begin{tabular}{|c|c|c|c|c|c|}
 \hline
\textbf{\begin{tabular}[c]{@{}c@{}}Accuracy\\ (RMS error)\end{tabular}}
 & 
 \textbf{\begin{tabular}[c]{@{}c@{}}Workload\\ (Multiplies)\end{tabular}}
 & 
 \textbf{\begin{tabular}[c]{@{}c@{}}\# LUTS\\  \end{tabular}}
 & 
 \textbf{\begin{tabular}[c]{@{}c@{}}\# DSPs\\  \end{tabular}}
 &
 \textbf{\begin{tabular}[c]{@{}c@{}}Latency ({\textmu}s)\\  \end{tabular}}
 
 & 
  \textbf{\begin{tabular}[c]{@{}c@{}}Optimized RF for Each Layer\\ \end{tabular}}
                                 \\ \hline
0.169                & 11.9K                & 18999                                    & 10                              & 168.83                          & 48, 768, 384, 768, 384, 64                                \\ \hline
0.1433               & 12.2K                & 24808                                    & 17                              & 169.14                          & 48, 384, 384, 384, 768, 64, 16, 16, 16, 4                 \\ \hline
0.1339               & 12.3K                & 24807                                    & 17                              & 169.14                          & 48, 768, 768, 384, 768, 64, 25, 25, 25, 5                 \\ \hline
0.119                & 12.6K                & 24807                                    & 17                              & 169.14                          & 48, 384, 768, 384, 768, 512, 32, 32, 32, 4                \\ \hline
0.1161               & 13.7K                & 26375                                    & 16                              & 171.82                          & 48, 768, 768, 768, 768, 384, 162, 162, 18                 \\ \hline
0.1134               & 15.7K                & 26375                                    & 16                              & 171.82                          & 48, 768, 768, 768, 768, 384, 162, 162, 18                 \\ \hline
0.1095               & 16.8K                & 27125                                    & 14                              & 171.82                          & 60, 600, 1200, 300, 1200, 1360, 289, 289, 17              \\ \hline
0.1065               & 21.7K                & 63052                                    & 40                              & 193.92                          & 78, 2028,  1014, 2028, 2028, 1768, 289, 289, 17           \\ \hline
0.1029               & 25.0K                 & 63052                                    & 40                              & 193.92                          & 90, 2700, 2700, 2700, 2700, 2040, 289, 289, 17            \\ \hline
0.0982               & 25.6K                & 30836                                    & 24                              & 170.59                          & 24, 192, 384, 768, 384, 1824, 1444, 38                    \\ \hline
0.0958               & 33.0K                 & 44702                                    & 30                              & 176.81                          & 24, 192, 384, 384, 768, 4512, 2209, 2209, 2209, 2209, 47  \\ \hline
0.0939               & 34.4K                & 63052                                    & 40                              & 194.94                          & 123, 5043, 5043, 5043, 5043, 3116, 361, 361, 19           \\ \hline
0.0851               & 36.6K                & 80227                                    & 58                              & 174.88                          & 24, 192, 768, 768, 384, 5600, 2500, 2500, 2500, 50        \\ \hline
0.0828               & 41.4K                & 91708                                    & 66                              & 176.96                          & 24, 192, 768, 768, 768, 336, 2916, 2916, 2916, 2916, 54   \\ \hline
0.0813               & 70.5K                & 91702                                    & 66                              & 176.96                          & 24, 192, 768, 768, 768, 13200, 5625, 5625, 5625, 5625, 75 \\ \hline
0.0792               & 74.9K                & 94960                                    & 78                              & 193.26                          & 24, 192, 192, 192, 768, 14592, 5776, 5776, 5776, 5776, 76 \\ \hline
\end{tabular}
\end{center}
\caption{Training and deployment results for Pareto optimal networks with $200~\mu\text{s}$ latency constraint, with corresponding estimated resource cost, estimated latency, and corresponding optimized reuse factor (RF) for each layer.}
\label{tab:pareto_front_networks}
\end{table*}

\subsection{Pareto Optimal Model Search}
\label{sec:pareto}

Using Optuna, \cite{optuna_2019} we obtain a set of Pareto optimal networks for the DROPBEAR dataset, using the dual objectives of accuracy and workload, as expressed by the number of multiplies required by one network inference. Table \ref{tab:pareto_front_networks} shows the Pareto optimal set of one such run.

We ran each of these networks through our MIP-based optimizer to set the reuse factor for each layer to ensure the whole network meets the latency constraint of $200~\mu\text{s}$ while minimizing its cost in LUTs, DSPs, registers, and BRAMs. For each network, the reuse factors for each layer are shown in the rightmost column. We expect the number of multiplies to correlate with resource cost, since given the fixed execution time, the number of multiplies relates directly to throughput. As shown in the table, this is generally true except for the networks in rows 8 and 9, whose LUT usage exceeds that of networks in rows 10 and 11 despite those networks having more multiplies. This is caused by imperfections in our resource cost model.


The networks in row 4 (model 1) and row 16 (model 2) are used to generate the results shown in Fig. \ref{fig:dropbear_accuracy} as ``model 2'' (shown in cyan) and ``model 1'' (shown in red), respectively. This plot shows a portion of one of the ``standard index sets'' of our \textit{Test Dataset 1} between 34.0 and 37.5 seconds, in which the roller traverses its full dynamic range two times.


\subsection{Cost and Latency Model Accuracy}

\begin{figure*}[ht]
    \centering
    \includegraphics[width=0.8\textwidth]{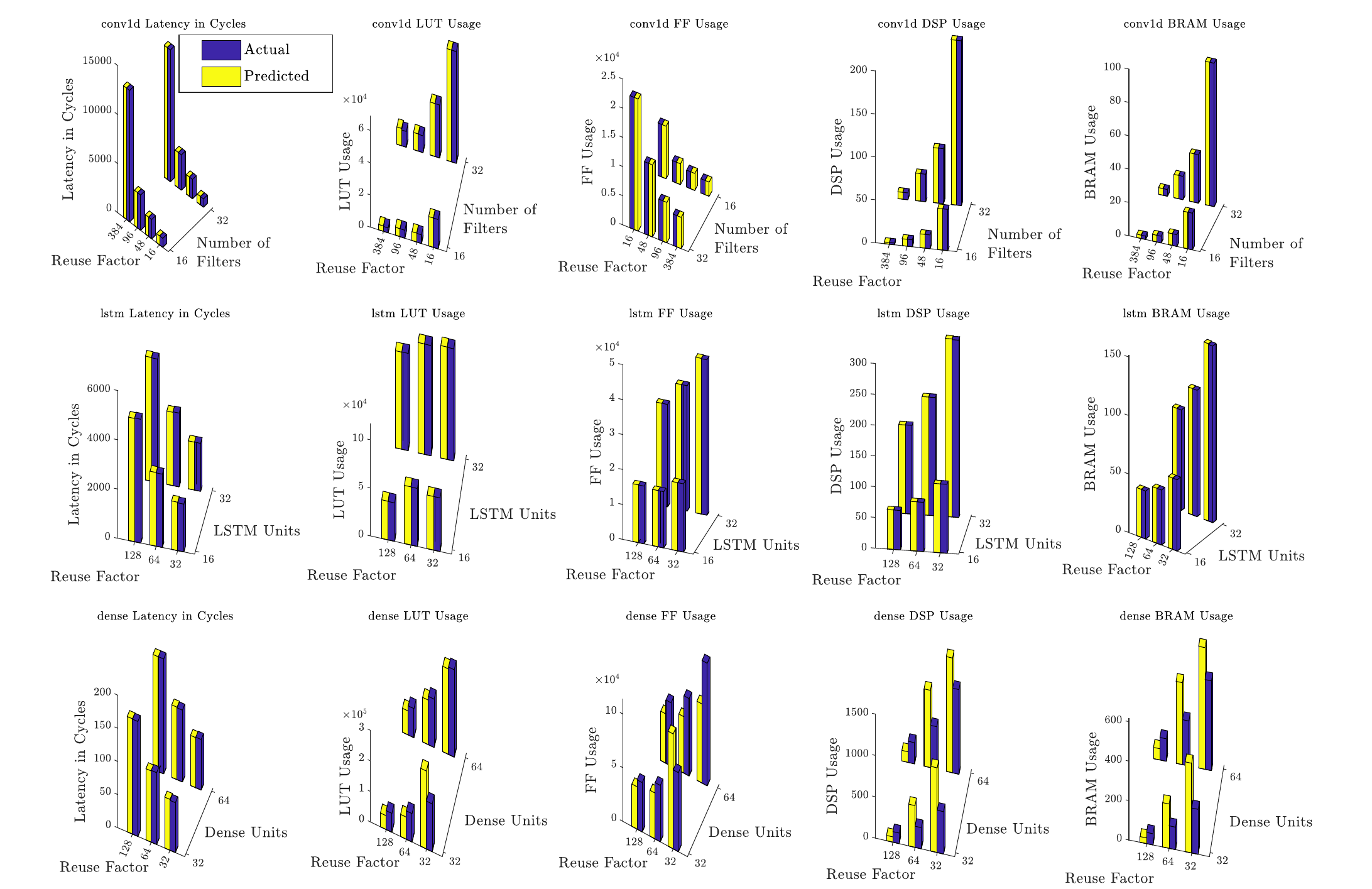}
    \caption{Performance and cost model results. conv1d results are for a layer having an input tensor size of (64,16), meaning a sequence length of 64 and embedding dimension of 16, LSTM results are for a layer having an input tensor size of (32,16), and dense results are for a layer having an input size of (1,512).}
    \label{fig:training_plots}
\end{figure*}

Fig.~\ref{fig:training_plots} presents a series of 3D bar plots illustrating (1) the correlation between latency and cost in relation to the reuse factor for various layer sizes with a fixed input tensor size, and (2) the corresponding prediction from our performance and cost models. The empirical values are derived from the HLS compiler (Vivado HLS 2019.1).


In these plots, the x-axis corresponds to the reuse factor while the y-axis corresponds to the size of each layer, which includes the number of filters for the 1D convolution layer, the number of units for the LSTM layer, and the number of neurons for the dense layer. The z-axis corresponds to the predicted latency or cost. The blue bar represents the ground truth values and the yellow bar represents the values predicted by our model. The input combinations shown in each figure are excluded from the training set used to generate the model being evaluated.

\begin{table*}[]
\centering
\resizebox{\textwidth}{!}{
\begin{tabular}{|l|l|cccc|cccc|cccc|}
\hline
\multirow{2}{*}{\textbf{Network}} &
  \multirow{2}{*}{\textbf{Trials}} &
  \multicolumn{4}{c|}{\textbf{Stochastic Search}} &
  \multicolumn{4}{c|}{\textbf{Simulated Annealing (SA)}} &
  \multicolumn{4}{c|}{\textbf{N-TORC}} \\
 &
   &
  \# LUTs &
  \# DSP &
  Latency ({\textmu}s) &
  Search Time (s) &
  \# LUTs &
  \# DSP &
  Latency ({\textmu}s) &
  Search Time (s) &
  \# LUTs &
  \# DSP &
  Latency ({\textmu}s) &
  Search Time (s) \\ \hline
\multirow{4}{*}{\begin{tabular}[c]{@{}l@{}}Model 1\\ \\ 1.3e11 RF\\ permuations\end{tabular}} &
  1K &
  137034 &
  209 &
  124 &
  5 &
  120481 &
  159 &
  111 &
  4 &
  94960 &
  78 &
  193 &
  5 \\
 &
  10K &
  106522 &
  134 &
  189 &
  47 &
  104306 &
  101 &
  162 &
  38 &
   &
   &
   &
   \\
 &
  100K &
  100054 &
  107 &
  140 &
  413 &
  98289 &
  101 &
  156 &
  382 &
   &
   &
   &
   \\
 &
  1M &
  95537 &
  79 &
  192 &
  4573 &
  93046 &
  136 &
  193 &
  3995 &
   &
   &
   &
   \\ \hline
\multirow{4}{*}{\begin{tabular}[c]{@{}l@{}}Model 2\\ \\ 3.4e11 RF\\ permuations\end{tabular}} &
  1K &
  445328 &
  746 &
  190 &
  6 &
  434219 &
  720 &
  162 &
  6 &
  374009 &
  459 &
  199 &
  6 \\
 &
  10K &
  415243 &
  646 &
  198 &
  53 &
  398131 &
  576 &
  196 &
  56 &
  \multicolumn{1}{l}{} &
  \multicolumn{1}{l}{} &
  \multicolumn{1}{l}{} &
  \multicolumn{1}{l|}{} \\
 &
  100K &
  391543 &
  508 &
  191 &
  565 &
  396019 &
  514 &
  187 &
  567 &
  \multicolumn{1}{l}{} &
  \multicolumn{1}{l}{} &
  \multicolumn{1}{l}{} &
  \multicolumn{1}{l|}{} \\
 &
  1M &
  383849 &
  474 &
  190 &
  5406 &
  376416 &
  466 &
  196 &
  4694 &
  \multicolumn{1}{l}{} &
  \multicolumn{1}{l}{} &
  \multicolumn{1}{l}{} &
  \multicolumn{1}{l|}{} \\ \hline
\end{tabular}
}
\caption{N-TORC Comparison with Stochastic Search and Simulated Annealing.}
\label{tab:search_comparison}
\end{table*}

Table ~\ref{tab:pareto_front_networks} shows the results from 16 DROPBEAR models. For these results, we consider two target networks optimized using the N-TORC framework with a real-time latency constraint of $200~\mu\text{s}$. The models are sorted in ascending order of accuracy, with the corresponding model workload and high-level synthesis results given from the reuse factor optimization of the network. As shown, each deployed network requires an end-to-end latency of slightly less than the constraint of $200~\mu\text{s}$ and the required number of LUTs and DSPs correlates with the workload. The effective throughput of the deployed models ranges from 11 Mops/s to 39 Mops/s and from 3.7\% to 18.8\% of the available LUTS and 0.58\% and 4.5\% of available DSPs on the XCZU7EV Zynq UltraScale+ FPGA. The last column shows the layer reuse factors assigned to achieve the corresponding results.

\subsection{Model Deployment Optimizer vs Stochastic Search}

To evaluate the execution time of N-TORC we compare it against a naive stochastic method and a simulated annealing method. For these results, we consider two target DROPBEAR models. Model 1 has 11 layers: 5 conv1d layers and 6 dense layers, and model 2 has 11 layers:  4 conv1d layers, 2 LSTM layers, and 5 dense layers.

The naive stochastic search method randomly assigns reuse factors to each layer and estimates the resultant resource cost and latency. After a given number of trials, it returns the assignment giving the minimum resource cost without exceeding the latency constraint.

The simulated annealing approach begins with a random reuse factor assignment for each layer and changes one each iteration. It accepts any assignment that gives the lowest-found resource cost while meeting the latency constraint, or any network that meets the latency constraint with probability $e^{\frac{r_{best} - r_{proposed}}{t}}$, with $t$ starting at 100 and cooling at a rate of 1\% per iteration and where $r_{best}$ is the resource cost of the best assignment found and $r_{proposed}$ is the resource cost of the most recent assignment.

Table \ref{tab:search_comparison} compares the execution time and optimization quality of the stochastic search, simulated annealing (SA) search, and N-TORC MIP approach. The results show stochastic and SA search runs having 1K, 10K, 100K, and 1M random trials. The stochastic and SA search time scales linearly.

To achieve comparable results, the stochastic and SA search require 1M trials, requiring 1000X the execution time of the MIP optimization. When compared to the result found with stochastic search and SA at 1M trials, N-TORC finds a configuration that requires significantly fewer DSPs for Model 1 and slightly fewer LUTS for Model 2.



\section{Related Work}

There are several recent efforts to estimate the energy, throughput, and resources required for HLS implementation of machine learning models. Xu et al. developed an analytical model that predicts the energy, latency, and resource utilization given the total workload associated with the DNN and the unrolling factor used (equivalent to the inverse of the HLS4ML reuse factor), and performs a cycle-accurate simulation of the design to gather more accurate estimates \cite{xu2020autodnnchip}. It is based on a reusable systolic array-based GEMM architecture. Because the systolic array is shared among all the layers, the framework cannot achieve fine-grain control of end-to-end latency, since changing the size of the systolic array has varying effects on the latency of different layers.

Shahshahani et al. propose an analytical model that predicts latency, memory overhead, and energy consumption for generated HLS code that performs 2D convolution, given the required workload and available memory throughput. The model considers the effects of loop pipelining and unrolling \cite{shahshahani2020framework}. Their approach includes an exhaustive design space exploration that optimizes an objective function that assigns priorities to latency and area, as opposed to performing a constrained optimization. 

Makrani et al. developed a data-driven model called Pyramid that re-calibrates the timing and resource results given by the HLS compiler to produce a more accurate estimate of the post-low-level synthesis results, but still requires high-level synthesis when performing design space exploration of high-level designs \cite{makrani2019pyramid}.


\section{Conclusion and Future Work}

In this paper we describe N-TORC, a tool flow for generating a candidate set of neural network models for a target dataset that achieves the highest possible accuracy for a given resource cost while meeting a latency constraint. We evaluate this approach using a benchmark structural state estimation dataset, DROPBEAR, but in principle, the approach can be used for any dataset that can be trained with a model whose parameters and output tensors can fit in on-chip memory of an embedded-class FPGA. The proposed method consists of two stages. The first stage searches the space of network configurations and hyperparameters to find a Pareto optimal set of trained models with respect to latency and workload, while the second stage searches for the optimal assignment of reuse factors to the model's layers to meet the end-to-end latency constraint while minimizing resource cost. The second stage is built on performance and cost models that predict resource cost and latency for CNN, LSTM, and dense layers, which we show have a prediction error of approximately 2\% over our test set. These performance models take advantage of being trainined specifically for the HLS4ML code structures, as opposed to being applicable to general-purpose HLS code \cite{wu2022high} or analytical models that assume the use of a generic systolic array circuit with a folding factor \cite{xu2020autodnnchip,shahshahani2020framework}.

A limitation of this work is that it does not consider network quantization, an increasingly common technique \cite {que2023metaml,9444059} that, if integrated into N-TORC, may further reduce the resource cost of deployed models. Since HLS4ML supports quantization in both weights and activations (in the current work we set both as 16-bit fixed point), we will incorporate quantization optimization into our future work.

\section*{Acknowledgments}

This material is based upon work supported by the National Science Foundation under Grant No. 1956071.



\bibliographystyle{IEEEtran}
%
\bibliography{references}

\begin{thebibliography}{10}
\providecommand{\url}[1]{#1}
\csname url@samestyle\endcsname
\providecommand{\newblock}{\relax}
\providecommand{\bibinfo}[2]{#2}
\providecommand{\BIBentrySTDinterwordspacing}{\spaceskip=0pt\relax}
\providecommand{\BIBentryALTinterwordstretchfactor}{4}
\providecommand{\BIBentryALTinterwordspacing}{\spaceskip=\fontdimen2\font plus
\BIBentryALTinterwordstretchfactor\fontdimen3\font minus \fontdimen4\font\relax}
\providecommand{\BIBforeignlanguage}[2]{{%
\expandafter\ifx\csname l@#1\endcsname\relax
\typeout{** WARNING: IEEEtran.bst: No hyphenation pattern has been}%
\typeout{** loaded for the language `#1'. Using the pattern for}%
\typeout{** the default language instead.}%
\else
\language=\csname l@#1\endcsname
\fi
#2}}
\providecommand{\BIBdecl}{\relax}
\BIBdecl

\bibitem{satme2022progress}
J.~Satme, D.~Coble, B.~Priddy, A.~R. Downey, J.~D. Bakos, and G.~Comert, ``Progress towards data-driven high-rate structural state estimation on edge computing devices,'' in \emph{International Design Engineering Technical Conferences and Computers and Information in Engineering Conference}, vol. 86311.\hskip 1em plus 0.5em minus 0.4em\relax American Society of Mechanical Engineers, 2022, p. V010T10A017.

\bibitem{kabir2023accelerating}
E.~Kabir, D.~Coble, J.~N. Satme, A.~R. Downey, J.~D. Bakos, D.~Andrews, and M.~Huang, ``Accelerating lstm-based high-rate dynamic system models,'' in \emph{2023 33rd International Conference on Field-Programmable Logic and Applications (FPL)}.\hskip 1em plus 0.5em minus 0.4em\relax IEEE, 2023, pp. 327--332.

\bibitem{vereen2023optimal}
A.~B. Vereen, E.~A. Ogunniyi, A.~R. Downey, E.~Blasch, J.~D. Bakos, and J.~Dodson, ``Optimal sampling methodologies for high-rate structural twinning,'' in \emph{2023 26th International Conference on Information Fusion (FUSION)}.\hskip 1em plus 0.5em minus 0.4em\relax IEEE, 2023, pp. 1--8.

\bibitem{ogunniyi2023real}
E.~A. Ogunniyi, C.~Drnek, S.~H. Hong, A.~R. Downey, Y.~Wang, J.~D. Bakos, P.~Avitabile, and J.~Dodson, ``Real-time structural model updating using local eigenvalue modification procedure for applications in high-rate dynamic events,'' \emph{Mechanical Systems and Signal Processing}, vol. 195, p. 110318, 2023.

\bibitem{Downey2021Dataset2Dropbear}
A.~Vereen, A.~Downey, J.~Dodson, and A.~G. Moura, ``Dataset-8-dropbear-acceleration-vs-roller-displacement,'' https://github.com/High-Rate-SHM-Working-Group/Dataset-8-DROPBEAR-Acceleration-vs-Roller-Displacement, Aug. 2023.

\bibitem{que2023metaml}
Z.~Que, S.~Liu, M.~Rognlien, C.~Guo, J.~G. Coutinho, and W.~Luk, ``Metaml: Automating customizable cross-stage design-flow for deep learning acceleration,'' in \emph{2023 33rd International Conference on Field-Programmable Logic and Applications (FPL)}.\hskip 1em plus 0.5em minus 0.4em\relax IEEE, 2023, pp. 248--252.

\bibitem{wielgosz2017using}
M.~Wielgosz, A.~Skocze{\'n}, and M.~Mertik, ``Using lstm recurrent neural networks for monitoring the lhc superconducting magnets,'' \emph{Nuclear Instruments and Methods in Physics Research Section A: Accelerators, Spectrometers, Detectors and Associated Equipment}, vol. 867, pp. 40--50, 2017.

\bibitem{egan2017long}
S.~Egan, W.~Fedorko, A.~Lister, J.~Pearkes, and C.~Gay, ``Long short-term memory (lstm) networks with jet constituents for boosted top tagging at the lhc,'' \emph{arXiv preprint arXiv:1711.09059}, 2017.

\bibitem{9221584}
Y.~Umuroglu, Y.~Akhauri, N.~J. Fraser, and M.~Blott, ``Logicnets: Co-designed neural networks and circuits for extreme-throughput applications,'' in \emph{2020 30th International Conference on Field-Programmable Logic and Applications (FPL)}, 2020, pp. 291--297.

\bibitem{YAN2021107960}
J.~Yan, S.~Laflamme, J.~Hong, and J.~Dodson, ``Online parameter estimation under non-persistent excitations for high-rate dynamic systems,'' \emph{Mechanical Systems and Signal Processing}, vol. 161, p. 107960, 2021.

\bibitem{vibration3030016}
\BIBentryALTinterwordspacing
J.~Yan, S.~Laflamme, P.~Singh, A.~Sadhu, and J.~Dodson, ``A comparison of time-frequency methods for real-time application to high-rate dynamic systems,'' \emph{Vibration}, vol.~3, no.~3, pp. 204--216, 2020. [Online]. Available: \url{https://www.mdpi.com/2571-631X/3/3/16}
\BIBentrySTDinterwordspacing

\bibitem{10.1117/12.3010900}
\BIBentryALTinterwordspacing
A.~Razmarashooli, D.~A.~S. Martinez, Y.~K. Chua, S.~Laflamme, and C.~Hu, ``{Real-time state estimation using recurrent neural network and topological data analysis},'' in \emph{Nondestructive Characterization and Monitoring of Advanced Materials, Aerospace, Civil Infrastructure, and Transportation XVIII}, A.~L. Gyekenyesi, P.~J. Shull, H.~F. Wu, and T.~Yu, Eds., vol. 12950, International Society for Optics and Photonics.\hskip 1em plus 0.5em minus 0.4em\relax SPIE, 2024, p. 129500C. [Online]. Available: \url{https://doi.org/10.1117/12.3010900}
\BIBentrySTDinterwordspacing

\bibitem{projectrepo}
S.~V. Singh, I.~Ahmad, D.~Andrews, M.~Huang, A.~R.~J. Downey, and J.~D. Bakos, ``N-torc: Native tensor optimizer for real-time constraints,'' https://github.com/HeRCLab/N-TORC, Aug. 2024.

\bibitem{Nelson2022GeneratedDatasetsDynamic}
M.~Nelson, S.~Laflamme, C.~Hu, A.~G. Moura, J.~Hong, A.~Downey, P.~Lander, Y.~Wang, E.~Blasch, and J.~Dodson, ``Generated datasets from dynamic reproduction of projectiles in ballistic environments for advanced research ({DROPBEAR}) testbed,'' \emph{{IOP} {SciNotes}}, vol.~3, no.~4, p. 044401, nov 2022.

\bibitem{Dodson2021HighRateStructural}
J.~Dodson, A.~Downey, S.~Laflamme, M.~D. Todd, A.~G. Moura, Y.~Wang, Z.~Mao, P.~Avitabile, and E.~Blasch, ``High-rate structural health monitoring and prognostics: An overview,'' in \emph{Data Science in Engineering, Volume 9}.\hskip 1em plus 0.5em minus 0.4em\relax Springer International Publishing, oct 2021, pp. 213--217.

\bibitem{panahi2022high}
A.~Panahi, E.~Kabir, A.~Downey, D.~Andrews, M.~Huang, and J.~D. Bakos, ``High-rate machine learning for forecasting time-series signals,'' in \emph{2022 IEEE 30th Annual International Symposium on Field-Programmable Custom Computing Machines (FCCM)}.\hskip 1em plus 0.5em minus 0.4em\relax IEEE, 2022, pp. 1--9.

\bibitem{hrwg}
A.~Downey and J.~Dodson, ``High-rate structural monitoring, damage detection, prognostics, and reactions working group,'' https://github.com/High-Rate-SHM-Working-Group, Aug. 2023.

\bibitem{10.1007/BFb0091924}
F.~Takens, ``Detecting strange attractors in turbulence,'' in \emph{Dynamical Systems and Turbulence, Warwick 1980}, D.~Rand and L.-S. Young, Eds.\hskip 1em plus 0.5em minus 0.4em\relax Berlin, Heidelberg: Springer Berlin Heidelberg, 1981, pp. 366--381.

\bibitem{10.1117/12.2613208}
\BIBentryALTinterwordspacing
E.~A. Ogunniyi, A.~R. J.~D. Jr., and J.~D. Bakos, ``{Development of a real-time solver for the local eigenvalue modification procedure},'' in \emph{Sensors and Smart Structures Technologies for Civil, Mechanical, and Aerospace Systems 2022}, D.~Zonta, B.~Glisic, and Z.~Su, Eds., vol. 12046, International Society for Optics and Photonics.\hskip 1em plus 0.5em minus 0.4em\relax SPIE, 2022, p. 120460U. [Online]. Available: \url{https://doi.org/10.1117/12.2613208}
\BIBentrySTDinterwordspacing

\bibitem{OGUNNIYI2023110318}
\BIBentryALTinterwordspacing
E.~A. Ogunniyi, C.~Drnek, S.~H. Hong, A.~R. Downey, Y.~Wang, J.~D. Bakos, P.~Avitabile, and J.~Dodson, ``Real-time structural model updating using local eigenvalue modification procedure for applications in high-rate dynamic events,'' \emph{Mechanical Systems and Signal Processing}, vol. 195, p. 110318, 2023. [Online]. Available: \url{https://www.sciencedirect.com/science/article/pii/S088832702300225X}
\BIBentrySTDinterwordspacing

\bibitem{10224187}
A.~B. Vereen, E.~A. Ogunniyi, A.~R. Downey, E.~Blasch, J.~D. Bakos, and J.~Dodson, ``Optimal sampling methodologies for high-rate structural twinning,'' in \emph{2023 26th International Conference on Information Fusion (FUSION)}, 2023, pp. 1--8.

\bibitem{duarte2018fast}
J.~Duarte, S.~Han, P.~Harris, S.~Jindariani, E.~Kreinar, B.~Kreis, J.~Ngadiuba, M.~Pierini, R.~Rivera, N.~Tran \emph{et~al.}, ``Fast inference of deep neural networks in fpgas for particle physics,'' \emph{Journal of instrumentation}, vol.~13, no.~07, p. P07027, 2018.

\bibitem{optuna_2019}
T.~Akiba, S.~Sano, T.~Yanase, T.~Ohta, and M.~Koyama, ``Optuna: A next-generation hyperparameter optimization framework,'' in \emph{Proceedings of the 25th {ACM} {SIGKDD} International Conference on Knowledge Discovery and Data Mining}, 2019.

\bibitem{balandat2020botorch}
\BIBentryALTinterwordspacing
M.~Balandat, B.~Karrer, D.~R. Jiang, S.~Daulton, B.~Letham, A.~G. Wilson, and E.~Bakshy, ``{BoTorch: A Framework for Efficient Monte-Carlo Bayesian Optimization},'' in \emph{Advances in Neural Information Processing Systems 33}, 2020. [Online]. Available: \url{http://arxiv.org/abs/1910.06403}
\BIBentrySTDinterwordspacing

\bibitem{scikit-learn}
F.~Pedregosa, G.~Varoquaux, A.~Gramfort, V.~Michel, B.~Thirion, O.~Grisel, M.~Blondel, P.~Prettenhofer, R.~Weiss, V.~Dubourg, J.~Vanderplas, A.~Passos, D.~Cournapeau, M.~Brucher, M.~Perrot, and E.~Duchesnay, ``Scikit-learn: Machine learning in {P}ython,'' \emph{Journal of Machine Learning Research}, vol.~12, pp. 2825--2830, 2011.

\bibitem{wu2022high}
N.~Wu, H.~Yang, Y.~Xie, P.~Li, and C.~Hao, ``High-level synthesis performance prediction using gnns: Benchmarking, modeling, and advancing,'' in \emph{Proceedings of the 59th ACM/IEEE Design Automation Conference}, 2022, pp. 49--54.

\bibitem{gurobi}
{Gurobi Optimization, LLC}, ``{Gurobi Optimizer Reference Manual},'' https://www.gurobi.com, 2024.

\bibitem{suyash}
\BIBentryALTinterwordspacing
S.~V. Singh, I.~Ahmad, D.~Andrews, M.~Huang, A.~R.~J. Downey, and J.~D. Bakos, ``Resource scheduling for real-time machine learning,'' in \emph{Proceedings of the 2025 ACM/SIGDA International Symposium on Field Programmable Gate Arrays}, ser. FPGA '25.\hskip 1em plus 0.5em minus 0.4em\relax New York, NY, USA: Association for Computing Machinery, 2025, p.~50. [Online]. Available: \url{https://doi.org/10.1145/3706628.3708848}
\BIBentrySTDinterwordspacing

\bibitem{xu2020autodnnchip}
P.~Xu, X.~Zhang, C.~Hao, Y.~Zhao, Y.~Zhang, Y.~Wang, C.~Li, Z.~Guan, D.~Chen, and Y.~Lin, ``Autodnnchip: An automated dnn chip predictor and builder for both fpgas and asics,'' in \emph{Proceedings of the 2020 ACM/SIGDA International Symposium on Field-Programmable Gate Arrays}, 2020, pp. 40--50.

\bibitem{shahshahani2020framework}
M.~Shahshahani, B.~Khabbazan, M.~Sabri, and D.~Bhatia, ``A framework for modeling, optimizing, and implementing dnns on fpga using hls,'' in \emph{2020 IEEE 14th Dallas Circuits and Systems Conference (DCAS)}.\hskip 1em plus 0.5em minus 0.4em\relax IEEE, 2020, pp. 1--6.

\bibitem{makrani2019pyramid}
H.~M. Makrani, F.~Farahmand, H.~Sayadi, S.~Bondi, S.~M.~P. Dinakarrao, H.~Homayoun, and S.~Rafatirad, ``Pyramid: Machine learning framework to estimate the optimal timing and resource usage of a high-level synthesis design,'' in \emph{2019 29th International Conference on Field Programmable Logic and Applications (FPL)}.\hskip 1em plus 0.5em minus 0.4em\relax IEEE, 2019, pp. 397--403.

\bibitem{9444059}
\BIBentryALTinterwordspacing
Z.~Dong, Y.~Gao, Q.~Huang, J.~Wawrzynek, H.~H. So, and K.~Keutzer, ``Hao: Hardware-aware neural architecture optimization for efficient inference,'' in \emph{2021 IEEE 29th Annual International Symposium on Field-Programmable Custom Computing Machines (FCCM)}.\hskip 1em plus 0.5em minus 0.4em\relax Los Alamitos, CA, USA: IEEE Computer Society, may 2021, pp. 50--59. [Online]. Available: \url{https://doi.ieeecomputersociety.org/10.1109/FCCM51124.2021.00014}
\BIBentrySTDinterwordspacing

\end{thebibliography}

\end{document}